\documentclass[sigconf, nonacm]{acmart}
\usepackage{graphicx} \usepackage{subcaption}
\usepackage{enumitem}
\usepackage{placeins}
\usepackage{multirow}

\title{Robust Fusion of Semantic and Behavioural Signals for LLM Reranking in Personalised Search}

\copyrightyear{2026}
\acmYear{2026}
\acmConference[USRW '26]{Unified Search and Recommendation Workshop}{October 2, 2026}{Minneapolis, MN, USA}

\author{Aleksandr V. Petrov}
\authornote{Authors contributed equally to this research.}
\email{aleksandrv@spotify.com}
\affiliation{\institution{Spotify}
\country{UK}
}

\author{Nathan Stein}
\authornotemark[1]
\email{nathanstein@spotify.com}
\affiliation{\institution{Spotify}
\country{Sweden}
}

\author{Erik Lybecker}
\authornotemark[1]
\email{erikl@spotify.com}
\affiliation{\institution{Spotify}
\country{Sweden}
}

\author{Emma Schüldt}
\email{eschuldt@spotify.com}
\affiliation{\institution{Spotify}
\country{Sweden}
}

\author{Daniel Lazarovski}
\email{dnll@spotify.com}
\affiliation{\institution{Spotify}
  \city{Stockholm}
  \country{Sweden}
}

\author{Hugues Bouchard}
\email{hb@spotify.com}
\affiliation{\institution{Spotify}
\country{Spain}
}

\author{Mounia Lalmas}
\email{mounia@acm.org}
\affiliation{\institution{Spotify}
\country{UK}
}



\begin{document}

\begin{abstract}

Personalised search sits at the boundary between retrieval and recommendation: it must satisfy explicit query intent while incorporating user context and collective interaction evidence. LLM-based cross-encoders offer a single reranking interface for these signals. In industrial search, however, relevance also depends heavily on behavioural signals such as click-through rates and historical search success. These signals can be injected directly into the language-model prompt, but doing so creates an important risk: the model may over-rely on historical statistics instead of learning semantic and user-context patterns that generalise to sparse or previously unseen searches.

We study this problem in the personalised search system of a large-scale audio streaming platform using Query Slice Stats (QSS), a collective, interaction-derived behavioural feature summarising historical success for query–candidate pairs. We first show that naïvely injecting QSS into the cross-encoder substantially improves ranking quality when QSS is available, but causes the model to depend heavily on this feature, reducing performance on searches where such statistics are sparse or missing. We address this problem with dual-sample feature-dropout training, a deterministic paired feature-removal strategy that presents each example once with QSS included and once with QSS intentionally removed.

Offline experiments show that injecting QSS improves ranking quality by 13.3\% when the feature is available; however, models trained with QSS always present perform worse in a diagnostic evaluation where QSS is removed. Dual-sample feature-dropout preserves these ranking gains while improving QSS-removed performance by 4.0\% relative to naïve QSS training.
In a live online test, both QSS-aware variants improve search success by roughly 2\%. The aggregate test does not distinguish dual-sample from features-only training, while the cold-start comparison is directionally consistent with the offline results.

These results show that paired feature-present and feature-removed training can reduce the tension between exploiting strong behavioural statistics and remaining robust when those statistics are unavailable.
\end{abstract}

\maketitle
\section{Introduction}
\label{sec:introduction}

Personalised search sits at the boundary between retrieval and recommendation: it combines explicit query intent with user preferences and collective evidence from historical interactions. On the one hand, it can exploit strong behavioural signals, such as historical click-through rates for frequent query-item pairs, which are often highly predictive on head queries. On the other hand, it must generalise to the long tail of rare and previously unseen queries and items, where such statistics are sparse or unavailable. This tension is amplified by the fact that the same query may require different rankings for different users, while query frequencies typically follow a heavy-tailed distribution with a small head and a vast tail of rare queries.

Pre-trained language models have recently transformed ranking in both information retrieval and recommender systems. In information retrieval, jointly encoding query and passage text with a pre-trained language model achieves state-of-the-art performance on benchmarks such as MS MARCO~\cite{nogueira2019passage,nguyen2016ms}. In recommender systems, jointly encoding user behaviour histories with LLMs similarly achieves state-of-the-art results on benchmarks such as MovieLens and Amazon Reviews~\cite{yue2023llamarec,harper2015movielens,hou2024bridging}. In personalised search, these two traditions meet within a single ranking problem: query and candidate content provide retrieval-style evidence, while user context and aggregate interactions provide recommendation-style evidence. Architectures that concatenate such conditioning context with each candidate and apply full self-attention across the pair are commonly termed \emph{cross-encoders}~\cite{urbanek2019learning,humeau2020poly,reimers2019sentence}.

Academic cross-encoders are typically evaluated on purely textual inputs, such as a query string paired with a document passage. Industrial ranking systems, however, operate in a much richer feature space. Production rankers rely heavily on numerical, categorical, and interaction-derived features, including click rates, success rates, impression counts, and popularity scores~\cite{karmaker2017application}. These signals are naturally represented as structured features rather than free text, and incorporating them into cross-encoders is non-trivial because the model input is primarily a sequence of tokens. Consequently, feature-based rankers such as gradient-boosted decision trees remain widely used industrial baselines~\cite{lutz2025industry}, partly because they handle heterogeneous feature types, skewed distributions, and missing values more naturally than standard neural architectures~\cite{grinsztajn2022tree}.

Recent LLM rerankers make it possible to expose such signals through the prompt by converting structured features into textual tokens. This provides a simple, production-compatible way to combine semantic, personalised, and behavioural signals within a single reranker without architectural modifications. However, we show that this flexibility introduces an important failure mode. When injected features are highly predictive of relevance, the model may over-rely on these behavioural statistics instead of learning semantic and personalised relevance patterns that generalise beyond frequent query-item pairs. As a result, ranking quality can degrade on sparse, long-tail, and cold-start traffic where historical statistics are unreliable or unavailable. We study this phenomenon through the lens of \emph{shortcut learning}~\cite{geirhos2020shortcut}, where models exploit easy predictive signals without learning the intended semantic behaviour.
The prompt therefore acts as a signal-fusion interface, combining semantic, user-context, and collective behavioural evidence.

Our work centres on Query Slice Stats (QSS), a collective, interaction-derived behavioural feature in the platform's search system that summarises historical success for query-item pairs.
QSS is particularly useful for studying this trade-off because it is highly predictive on frequent traffic while becoming sparse or unavailable for rare queries and newly introduced items.
We convert QSS values into ordinal prompt tokens, allowing the cross-encoder to condition directly on historical search success through the prompt.
We then apply dual-sample feature-dropout training, which pairs each QSS-present training example with a matched QSS-removed view.
This encourages the reranker to use behavioural statistics when informative while remaining effective when it must rely more heavily on semantic and user-context signals.

To evaluate this trade-off, we measure ranking quality both when QSS is available and when it is removed at evaluation time, providing a direct probe of the model's dependence on behavioural statistics. We further slice the analysis by query and item training frequency to distinguish behaviour on head traffic from sparse and cold-start regimes. Our experiments show that naïvely injecting QSS substantially improves ranking quality, but also causes the model to depend heavily on this feature. Dual-sample feature-dropout training preserves most of these gains while improving QSS-removed performance; popularity-sliced results show its clearest advantage on sparse and cold-start traffic.

In summary, our contributions are as follows:
\begin{enumerate}[leftmargin=*]
\item We describe a production-compatible prompt representation for injecting structured statistical features into LLM-based cross-encoders, instantiated through QSS in a large-scale personalised search system.
\item We identify shortcut learning as a practical failure mode of prompt-based feature injection: strong behavioural statistics improve ranking quality on frequent traffic, but can reduce performance on sparse and cold-start searches where such signals are unavailable.
\item We evaluate a deterministic paired feature-removal strategy that preserves QSS-derived ranking gains offline while improving QSS-removed and long-tail performance; a live A/B/C test separately confirms the production value of QSS-aware reranking.
\end{enumerate}

\section{Prior Work and Context}
\label{sec:context}

\paragraph{Cross-encoder reranking.}
Cross-encoders are established second-stage rerankers in information retrieval~\cite{nogueira2019passage}, with text-to-text, LLM-based, latency-aware, and deployed variants extending the architecture beyond academic benchmarks~\cite{nogueira2020document,ma2024fine,zhang2025qwen3,petrov2024shallow,puthenputhussery2025large}. We use a discriminative cross-encoder to score candidates from a first-stage retrieval system, following the candidate-generation-then-ranking pattern common to large-scale search and recommendation~\cite{covington2016deep,yue2023llamarec}. The distinctive question in our setting is how that reranker combines semantic, user-context, and interaction-derived evidence.

\paragraph{Semantic and behavioural signal fusion.}
Production rankers routinely combine memorisation-oriented sparse signals with learned representations that support generalisation~\cite{cheng2016wide,wang2021dcn}. Recent CTR and recommendation models extend this pattern by fusing language-model representations with non-textual features, collaborative embeddings, or numerical and categorical side information~\cite{wang2023bert4ctr,li2025ctrl,zhang2025collm,chen2024elcorec}. Behavioural aggregates over query--item histories likewise remain central in product-search ranking~\cite{liu2024long}. A close information-retrieval precedent injects a first-stage BM25 score into a BERT reranker as text~\cite{askari2023injecting}; our setting instead examines an interaction-derived query--candidate statistic and treats its prompt injection as a robustness problem when the statistic is sparse or unavailable.

\paragraph{Missing-signal robustness and cold start.}
Shortcut learning describes reliance on an easy predictive signal that does not capture the intended behaviour~\cite{geirhos2020shortcut}. Relevant methodological analogues deliberately train for degraded inputs: DropoutNet masks collaborative representations to support cold-start recommendation, while robust missing-input training exposes sequential recommenders to incomplete interaction histories~\cite{volkovs2017dropoutnet,siciliano2024robust}. Industrial retrieval work also shows that robust training objectives can improve generalisation, including for cold-start users, although it does not study missing prompt features directly~\cite{kolodner2024robust}. These approaches motivate our paired QSS-present and QSS-removed views, whereas work on semantic-ID generalisation, long-tail product search, and cold-item embeddings motivates the sparse-regime evaluation slices~\cite{singh2024better,kekuda2024embedding,zhu2021learning}. Our focus is whether a strong behavioural feature becomes a shortcut inside an LLM reranker, and whether deterministic paired feature removal reduces that reliance.

\section{System and Problem Setup}
\label{sec:setup}

This section describes the personalised search reranking setting studied in this paper. 
Section~\ref{sec:model} introduces the cross-encoder reranker and the production search candidates it reranks, while Section~\ref{sec:prompt-composition} describes how semantic, personalised, and behavioural signals are represented in the prompt.

\subsection{Model}
\label{sec:model}

Our reranker is built on a pre-trained open-source language model with 0.6B parameters. The model vocabulary is extended with Semantic IDs, grounded in platform interaction data using a procedure similar to~\cite{he2026plum,d2026deploying}. The reranker operates over candidates retrieved by the platform's search stack, which spans music, podcasts, and audiobooks.

Ranking is formulated as a pointwise prediction problem. For each query–candidate pair, the model jointly encodes the user context, search query, and candidate item, and produces a relevance score used to reorder the candidate set.

Personalisation is central to this setting. In personalised search, relevance is not determined by the query and candidate alone: user context provides additional evidence about intent and preference. The same query may correspond to different intents for different users, while recent and long-term behaviour, market, language, and other contextual signals can help disambiguate those intents. The cross-encoder therefore combines explicit intent from the query with implicit preference information derived from user context.

This formulation also reflects a broader shift in industrial ranking systems. Traditional production rankers typically combine heterogeneous semantic, behavioural, and contextual signals through specialised feature-engineering pipelines. In contrast, LLM-based rerankers provide a unified modelling framework in which these signals can be represented directly within the prompt and jointly processed by a single model.

\subsection{Prompt Composition}
\label{sec:prompt-composition}

Each scoring instance is represented as a structured prompt containing semantic, personalised, and behavioural information. The prompt includes the search query, user-context fields such as market and recent listening history, and candidate fields such as title, creator, content type, and a learned Semantic ID.
Semantic IDs provide compact discrete representations for catalogue entities, allowing item identity information to be encoded alongside natural-language metadata.

Beyond semantic and contextual information, the prompt also incorporates numerical and aggregate behavioural features. Before insertion, continuous or high-cardinality values are discretised into coarse textual tiers. Item popularity and query-item success statistics are therefore represented as bucketed feature tokens rather than raw counts or rates. This allows behavioural signals such as QSS (defined in Section~\ref{sec:qss}) to influence the reranker through the same textual interface as semantic signals.

This prompt design provides a unified textual interface through which the cross-encoder can jointly model semantic content, user context, and behavioural statistics. 
Figure~\ref{fig:prompt-schema} illustrates the main signal groups included in the prompt. The remainder of the paper focuses on QSS, the behavioural feature at the centre of our analysis, because it is both highly predictive on frequent query-item slices and potentially unreliable or unavailable in sparse regimes.

\begin{figure}[t]
    \centering
    \fbox{\begin{minipage}{0.92\linewidth}
    \footnotesize
\textbf{Semantic evidence:} query text; candidate title, creator, content type, and Semantic ID\\
    \textbf{User-context evidence:} country and recent Semantic IDs\\
    \textbf{Behavioural evidence:} popularity tier and QSS success-rate tier\\
    \textbf{Decision:} predict \texttt{Y} if the candidate satisfies the search
    \end{minipage}}
\caption{Schematic prompt composition for the cross-encoder, grouped into semantic, user-context, and behavioural evidence. The production prompt contains additional fields and formatting.}
    \Description{A boxed schematic prompt with user context, query, candidate, feature, and decision fields.}
    \label{fig:prompt-schema}
\end{figure}

\section{QSS-Aware Cross-Encoder Training}
\label{sec:methodology}

This section introduces the behavioural feature and training framework used to study shortcut learning in LLM-based rerankers. 
Section~\ref{sec:qss} defines Query Slice Stats (QSS), the historical query-item signal exposed to the cross-encoder, and explains why such features create a tension between memorisation and semantic generalisation. Section~\ref{sec:finetuning} then describes the supervised fine-tuning setup and the dual-sample training objective used to control reliance on behavioural statistics.

\subsection{Query Slice Stats (QSS)}
\label{sec:qss}

\paragraph{Feature definition.}

Query Slice Stats (QSS) is a production behavioural feature that captures historical search success for a query–candidate slice. The underlying feature store maintains aggregate interaction statistics for multiple slice definitions, including candidate item, query, query–item, and query–item–country combinations. Each slice includes counts together with click and search-success rates represented in basis points  (0--10{,}000).

In this work, QSS refers specifically to the query–item–country success-rate feature: the historical fraction of searches for the same query, candidate item, and user country that resulted in a successful interaction. This makes QSS a strong memorisation-style signal for frequently observed query-item combinations, while providing little or no useful evidence for rare or previously unseen ones.

Rather than injecting raw numerical values into the prompt, the continuous success rate is discretised into ordinal tiers derived from percentiles of the historical feature distribution: \texttt{very\_low}, \texttt{low}, \texttt{medium}, \texttt{high}, and \texttt{very\_high}.
Missing values are treated as \texttt{unknown} and omitted from the prompt. When present, the resulting tier is rendered as a candidate-level \texttt{success\_rate} token, allowing the cross-encoder to condition on behavioural evidence through the same textual interface used for semantic and contextual features.

\paragraph{Shortcut learning.}

Behavioural statistics such as QSS are often highly predictive on frequent queries because they encode repeated historical interactions. However, for rare or previously unseen query-item pairs, these signals may be sparse, noisy, or entirely unavailable. This issue is particularly important in personalised search, where newly introduced items and long-tail queries frequently lack sufficient historical evidence.

In these sparse regimes, the reranker must rely more heavily on semantic matching and user context. If training encourages excessive reliance on behavioural statistics, the model may exhibit shortcut learning: highly predictive historical signals can dominate semantic and personalised evidence needed for generalisation. The central challenge is therefore not simply how to expose behavioural statistics to an LLM reranker, but how to use them without degrading performance on sparse and cold-start traffic.
Our objective is to train the cross-encoder to exploit QSS when reliable while remaining effective when such signals are sparse or absent.

\subsection{Supervised Fine Tuning}
\label{sec:finetuning}

\paragraph{Pointwise relevance scoring.}

Given a search query and a candidate item, reranking is formulated as a binary relevance prediction problem. The prompt presents the user context, query, and candidate features, and the causal language model predicts a single relevance token: ``Y'' (relevant) or ``N'' (not relevant). At inference time, the log-probability assigned to ``Y'' at the decision position is used as the relevance score for reranking.

This label-token formulation follows the same principle as monoT5-style reranking~\cite{nogueira2020document}, where relevance prediction is cast as next-token generation over a small output vocabulary. The approach allows the reranker to leverage the pretrained language model directly, without introducing task-specific scoring heads.

\paragraph{Supervised fine-tuning loss.}

Training uses standard cross-entropy loss. Given a prompt $x$, let $\mathbf{s} = (s_1, \ldots, s_T)$ denote the tokenised sequence and $y$ the target completion token. The supervised fine-tuning (SFT) training objective is:
\begin{align}
\mathcal{L}_{\text{SFT}}(x, y) = -\log p_{\theta}(y \mid \mathbf{s})
\end{align}    
where $\theta$ denotes model parameters. In our setting, $y \in \{\text{Y}, \text{N}\}$ is a single relevance label, and loss is computed only at the label position rather than across the full prompt sequence.

\paragraph{Dual-sample feature-dropout loss.}

Throughout, we use \emph{QSS-removed} to mean that QSS is intentionally removed from the prompt for training or evaluation, and \emph{QSS unavailable} to mean that reliable QSS is naturally missing or unreliable in sparse and cold-start regimes.

To discourage shortcut learning, we use a deterministic paired feature-removal strategy, which we call dual-sample feature-dropout training. This is a prompt-level data-augmentation scheme rather than a new learning principle: every labelled example contributes a QSS-present view and a matched QSS-removed view. The two prompts are:
\begin{itemize}
    \item $x^{+}$: the full prompt containing all features;
    \item $x^{-}$: a QSS-removed prompt identical to $x^{+}$ except that the QSS feature is intentionally removed.
\end{itemize}
Both prompts share the same target label $y$.
The loss is computed independently for each view and combined as
\begin{align}
     \mathcal{L}_{\text{mixture}} =
     \alpha \cdot \mathcal{L}_{\text{SFT}}(x^{+}, y)
     + (1 - \alpha) \cdot \mathcal{L}_{\text{SFT}}(x^{-}, y)
\end{align}
where $\alpha \in [0, 1]$ controls the relative weight assigned to the QSS-present view.

Unlike ordinary training with the full prompt only, this objective explicitly exposes the model to both behavioural-feature-present and behavioural-feature-removed views of every labelled query--candidate pair.
The QSS-removed term therefore acts as a regulariser against excessive dependence on behavioural statistics, encouraging the reranker to remain effective when such signals are sparse, noisy, or unavailable.

Evaluating both views requires additional training computation relative to a single-view update. At serving time, however, the method uses the same single cross-encoder forward pass and unchanged prompt interface as the corresponding QSS-aware model.

\paragraph{Relation to random QSS dropout.}
The QSS-removed term can be viewed as a targeted prompt-level data augmentation strategy.
For each QSS-present example $x^{+}$, we construct a matched QSS-removed variant $x^{-}$ while keeping the target label $y$ unchanged.
This is similar in spirit to feature-level dropout, since the model is trained to make the correct prediction when a strongly predictive prompt feature is removed.

Random QSS dropout is a stochastic variant of the same feature-removal idea: rather than evaluating both views for every labelled example, it samples one view per update. At the level of expected QSS exposure, the dropout probability $p$ corresponds to the QSS-removed weight $1-\alpha$ in the dual-sample mixture. We evaluated random dropout in a separate offline sweep; it shows that stochastic feature removal can also improve QSS-removed performance, but it does not provide a matched statistical comparison establishing that either schedule is superior. We therefore treat random dropout as a supporting ablation and focus on the deterministic formulation because it makes the paired QSS-present/QSS-removed objective explicit.

This should not be confused with a standard neural \texttt{Dropout} layer, which randomly masks hidden activations inside the network during the forward pass.
Here, dropout refers to removing a semantically defined prompt feature before tokenisation.
The QSS-removed view must therefore be explicitly constructed by deleting the QSS feature from the input prompt.

Because the paired prompts differ only by the QSS feature, the deterministic formulation may also enable shared-prefix computation, for example through key-value caching in implementations that support it.

In summary, QSS is exposed to the reranker as a behavioural prompt feature, while the training objective controls the model's reliance on that feature through paired QSS-present and QSS-removed supervision.
The following section evaluates whether this design preserves the ranking gains from behavioural statistics while improving robustness in sparse and cold-start regimes.

\section{Experiments}
\label{sec:experiments}

We evaluate how behavioural statistics affect ranking quality, semantic generalisation, and robustness in LLM rerankers. We first establish the strength of QSS as a behavioural ranking signal and examine how injecting it into the prompt affects ranking performance. We then study the trade-off between exploiting behavioural statistics and preserving semantic generalisation under the dual-sample feature-dropout objective. Finally, we analyse how these effects vary across the query and item popularity spectrum, with particular focus on sparse and cold-start regimes.

\subsection{Experimental Setup}

\paragraph{Dataset.}
\begin{table}[t]
\caption{Dataset statistics. Training frequency is the number of
sessions containing a given query (or item) in the training set.
Long tail / short head split at frequency 500.}
\label{tab:dataset-stats}
\centering
\footnotesize
\begin{tabular}{lrr}
\toprule
& \textbf{Training} & \textbf{Eval} \\
\midrule
Sessions & 1,000,000 & 10,000 \\
Unique queries & 452,124 & 8,875 \\
Total candidates & 517,311,726 & 5,140,399 \\
Avg.\ candidates / session & 517 & 514 \\
Unique items & 46,149,749 & 2,519,719 \\
Positive items & 1,303,043 & 12,996 \\
Avg.\ positives / session & 1.30 & 1.30 \\
\midrule
\multicolumn{3}{l}{\textit{Eval sessions by query training frequency}} \\
\midrule
\quad 0 (cold start) & --- & 3,471 \\
\quad 1--5 & --- & 2,356 \\
\quad 6--20 & --- & 1,587 \\
\quad 21--100 & --- & 1,518 \\
\quad 101--500 & --- & 848 \\
\quad $>$500 & --- & 220 \\
\quad \textbf{Long tail ($\leq$500)} & --- & \textbf{9,780} \\
\quad \textbf{Short head ($>$500)} & --- & \textbf{220} \\
\midrule
\multicolumn{3}{l}{\textit{Eval sessions by avg.\ positive-item training frequency}} \\
\midrule
\quad 0 (cold start) & --- & 2,803 \\
\quad 1--5 & --- & 2,624 \\
\quad 6--20 & --- & 1,961 \\
\quad 21--100 & --- & 1,914 \\
\quad 101--500 & --- & 568 \\
\quad $>$500 & --- & 130 \\
\quad \textbf{Long tail ($\leq$500)} & --- & \textbf{9,870} \\
\quad \textbf{Short head ($>$500)} & --- & \textbf{130} \\
\bottomrule
\end{tabular}
\end{table} \label{sec:dataset}

Training and evaluation data are constructed from production search interaction logs collected over a two-week period at a large-scale audio streaming platform. Each session corresponds to a user-issued search query paired with the candidate set returned by the first-stage retrieval system. A candidate is labelled positive if the user has a successful interaction with it, such as streaming a track, following an artist, or adding a podcast to their library.
Table~\ref{tab:dataset-stats} summarises the dataset statistics.

The training set contains 1M sessions spanning 452K unique queries and 46.1M unique candidate items. The held-out evaluation set contains 10K sessions (8.9K unique queries and 2.5M unique items) sampled from a non-overlapping time period. Candidates span multiple content types, including tracks, artists, albums, playlists, podcast shows, podcast episodes, and audiobooks.

The task combines retrieval and recommendation characteristics. While the query provides an explicit intent signal, successful interactions depend not only on textual relevance but also on user preference. For example, a user searching for “workout playlist” may prefer a result aligned with their listening profile rather than the most lexically similar candidate. The reranker must therefore combine semantic query–candidate matching with personalised preference modelling.

The experimental setup is heavily sparse: sessions contain on average 514 retrieved candidates but only 1.3 positive items. This sparsity is central to the shortcut-learning problem studied in this paper. Behavioural statistics such as QSS are highly informative on frequent traffic, but sparse or unavailable on rare queries and cold-start items.

\paragraph{Popularity distribution.}

Query and item popularity follow a heavy-tailed distribution, which directly shapes the shortcut-learning behaviour studied in this paper. In the evaluation set, 34.7\% of queries are unseen during training (frequency 0), and the median query appears only 3 times in the training data. Only 2.2\% of evaluation sessions (220 out of 10K) belong to the short head, defined as queries appearing more than 500 times during training; the remaining 97.8\% form the long tail. A similar pattern holds for items: the median positive item appears 4 times in training, and only 1.6\% of sessions contain positive items from the short head.

Throughout the paper, we use a threshold of 500 to separate head from tail traffic. This boundary lies between the 95th and 99th percentiles of both query and item frequency distributions and corresponds to the transition between slices with rich behavioural evidence and sparse regimes where the model must rely more heavily on semantic generalisation. The same threshold is used consistently across all analyses and figures.\footnote{For simplicity, we define cold-start queries as queries with zero training occurrences and cold-start items as items with zero positive training occurrences. Unlike cold-start queries, cold-start items may still appear as non-relevant candidates in the training data.}

\paragraph{Metrics.}
\label{sec:metrics}

Ranking quality is evaluated using NDCG@7 and Recall@7, computed per session and averaged over the evaluation set.
NDCG@7 measures ranking quality while accounting for the positions of relevant results, whereas Recall@7 measures whether relevant items appear within the top-ranked results.
The cutoff of 7 corresponds to the number of results visible without scrolling in the production search interface, making it a natural production-relevant threshold.
Statistical significance is assessed using paired $t$-tests ($p<.05$) over session-level scores.

\paragraph{Training.}
\label{sec:training}

All models initialise from the same PLUM-style~\cite{he2026plum,d2026deploying} 0.6B-parameter cross-encoder checkpoint and are fine-tuned for one epoch on the 1M-session training set. Ranking is trained pointwise, with each query--candidate pair treated as an independent training instance. Optimisation uses a batch size of 4, a learning rate of $10^{-4}$ with a warmup-stable-decay schedule, and gradient clipping at~1.0.
To study the effect of dual-sample feature-dropout training, we vary the mixture weight $\alpha \in \{0,0.25,0.5,0.75,1.0\}$.
Each model is evaluated under two conditions: with the full production prompt including QSS (``All feat.''), and under QSS-removed evaluation, where QSS is intentionally removed from the prompt (``No QSS'').
The latter directly probes semantic generalisation independent of behavioural statistics.

\paragraph{Diagnostic reference points.}

Table~\ref{tab:offline-results} reports four diagnostic reference points evaluated on the same candidate sets as the cross-encoder variants. Random assigns candidates uniformly at random within each session. BM25 metadata performs lexical matching between the search query and candidate metadata using standard Okapi BM25 parameters ($k_1=1.2$, $b=0.75$) without tuning~\cite{robertson1994okapi}.
monoT5 applies text-to-text reranking to the search query and candidate metadata following the standard monoT5 setup in~\cite{nogueira2020document}.\footnote{We use the \texttt{castorini/monot5-base-msmarco-10k} checkpoint.}
BM25 metadata and monoT5 serve as lexical and pretrained text-only relevance anchors; neither is tuned or presented as a competitive production baseline.
QSS only scores candidates using the raw query--item success statistic before discretisation into prompt tiers and assigns missing values a score of 0. It measures how much standalone ranking signal the potential shortcut carries, rather than representing a complete ranking system.

\subsection{Results}
\begin{table*}[t]
\centering
\caption{Offline results on the 10K evaluation sessions.}
\label{tab:offline-results}
\footnotesize
\begin{tabular}{llcccc}
\toprule
& & \multicolumn{2}{c}{NDCG@7} & \multicolumn{2}{c}{Recall@7} \\
\cmidrule(lr){3-4} \cmidrule(lr){5-6}
Category & Method / config & All feat. & No QSS & All feat. & No QSS \\
\midrule
\multirow{4}{*}{Diagnostic anchors}
& Random & 0.010 & -- & 0.018 & -- \\
& BM25 metadata & 0.105 & -- & 0.160 & -- \\
& monoT5 & 0.104 & -- & 0.161 & -- \\
& QSS only & 0.494 & -- & 0.601 & -- \\
\midrule
\multirow{5}{*}{Cross-encoder variants}
& Dual-sample, $\alpha{=}0$$^{\S}$ 
& 0.549$^{\dagger}$ 
& \textbf{0.550}$^{\dagger}$ 
& 0.685$^{\dagger}$ 
& \textbf{0.686}$^{\dagger}$ \\
& Dual-sample, $\alpha{=}0.25$ 
& 0.618$^{*}$ 
& 0.547$^{\dagger}$ 
& 0.750$^{*}$ 
& 0.684$^{\dagger}$ \\
& Dual-sample, $\alpha{=}0.5$ 
& \textbf{0.622}$^{*\dagger\ddagger}$ 
& 0.547$^{\dagger}$ 
& \textbf{0.756}$^{*\dagger\ddagger}$ 
& 0.684$^{\dagger}$ \\
& Dual-sample, $\alpha{=}0.75$ 
& 0.616$^{*}$ 
& 0.539$^{*\dagger}$ 
& 0.753$^{*}$ 
& 0.674$^{*\dagger}$ \\
& Dual-sample, $\alpha{=}1$ 
& 0.618$^{*}$ 
& 0.526$^{*}$ 
& 0.750$^{*}$ 
& 0.658$^{*}$ \\
\bottomrule
\end{tabular}

\vspace{2pt}
\begin{minipage}{0.88\linewidth}
\scriptsize
\textit{Notes.}
``All feat.'' uses the full production prompt; ``No QSS'' removes QSS at evaluation time.
For dual-sample rows, $^{*}$ denotes sig.\ diff.\ vs $\alpha{=}0$ and $^{\dagger}$ denotes sig.\ diff.\ vs $\alpha{=}1$.
$^{\ddagger}$ denotes sig.\ diff.\ between the best dual-sample model and QSS only.
$^{\S}$Trained without QSS; unseen QSS tokens act as noise in the ``All feat.'' condition.
All tests use paired $t$-tests with $p<.05$.
\end{minipage}
\end{table*}

A central question in this work is whether behavioural statistics are strong enough to dominate semantic ranking signals during training.
Table~\ref{tab:offline-results} shows that this is indeed plausible.
BM25 metadata and monoT5 perform similarly, reaching NDCG@7 values of 0.105 and 0.104, respectively, with no statistically significant difference under paired $t$-tests over session-level scores ($p=0.43$).
In contrast, QSS only reaches 0.494 NDCG@7 and 0.601 Recall@7, significantly exceeding both text-only diagnostic anchors ($p<.001$).
However, QSS alone remains significantly below the best dual-sample model on NDCG@7: $\alpha{=}0.5$ reaches 0.622 with all features, a +0.128 absolute gain over QSS only.
This demonstrates that behavioural statistics alone capture substantial ranking signal, while effective cross-encoder ranking still benefits from semantic, personalised, and contextual evidence beyond this single behavioural feature.
The QSS-only diagnostic therefore serves as a direct measure of the potential shortcut's standalone strength: because QSS is highly predictive on its own, the model may learn to rely on it disproportionately when exposed to it through the prompt.

\subsubsection{Effect of QSS on ranking quality.}

To isolate the effect of behavioural statistics, we compare the baseline cross-encoder trained without QSS ($\alpha{=}0$) against the features-only variant trained with QSS always present ($\alpha{=}1.0$) in Table~\ref{tab:offline-results}.
When evaluated with the full production prompt (``All feat.''), adding QSS improves NDCG@7  by +12.6\% and Recall@7 by +9.5\%. Both improvements are statistically significant ($p<.05$, paired $t$-test), confirming that QSS provides a strong behavioural ranking signal.

However, these gains come at the expense of semantic generalisation. When the same $\alpha{=}1.0$ model is evaluated under the QSS-removed condition (``No QSS'' in Table~\ref{tab:offline-results}), its NDCG@7 falls to 0.526, significantly below the $\alpha{=}0$ baseline (0.550).
This indicates that the model has learned to rely heavily on QSS rather than on semantic and personalised relevance signals.

These results establish the central trade-off studied in this paper: behavioural statistics substantially improve ranking quality when available, but reliance on them degrades performance in sparse regimes where such signals are missing. We next examine whether dual-sample feature-dropout training can preserve the ranking gains from QSS while reducing this shortcut-learning behaviour.

\subsubsection{Trade-off between QSS utilisation and semantic generalisation.}

\begin{figure}[t]
    \centering
    \includegraphics[width=0.7\linewidth]{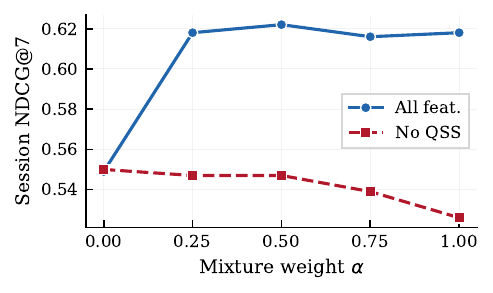}
    \caption{Offline NDCG@7 as the dual-sample mixture weight $\alpha$ varies. The ``No QSS'' curve probes semantic generalisation with QSS removed from the prompt.}
    \Description{NDCG@7 for different values of the dual-sample mixture weight alpha.}
    \label{fig:alpha-sweep-ndcg}
\end{figure}

To study how behavioural-feature exposure affects the balance between ranking quality and semantic generalisation, we sweep the dual-sample mixture weight $\alpha \in \{0, 0.25, 0.5, 0.75, 1.0\}$ and evaluate each model under full-prompt and QSS-removed evaluation at inference time  (Table~\ref{tab:offline-results} and Figure~\ref{fig:alpha-sweep-ndcg}).

The ``All feat.'' curve in Figure~\ref{fig:alpha-sweep-ndcg} shows that ranking quality improves rapidly even with limited QSS exposure during training. NDCG@7 increases from 0.549 at $\alpha{=}0$ to 0.618 at $\alpha{=}0.25$, already recovering most of the QSS-derived gain, and reaches 0.622 at $\alpha{=}0.5$ (+13.3\% over baseline). This indicates that only modest exposure to behavioural statistics is needed for the model to exploit their ranking signal.

The QSS-removed evaluation reveals the complementary effect on semantic generalisation.
At $\alpha{=}0.5$, NDCG@7 under QSS-removed evaluation remains statistically indistinguishable from the model trained without QSS (0.547 vs.\ 0.550; $p=0.15$), whereas $\alpha{=}1.0$ reduces performance significantly to 0.526.
The same pattern holds for Recall@7: $\alpha{=}0.5$ achieves the highest recall with all features (0.756) while preserving near-baseline recall under QSS-removed evaluation (0.684 vs.\ 0.686); see Table~\ref{tab:offline-results}.

Together, these results show that moderate QSS exposure captures the behavioural ranking gains while preserving semantic generalisation, whereas training with QSS always present encourages shortcut reliance and weakens performance under QSS-removed evaluation.

Overall, $\alpha{=}0.5$ provides the best trade-off between exploiting behavioural statistics and preserving semantic generalisation. We next examine how these gains are distributed across the query and item popularity spectrum.

\subsubsection{Distribution of improvements across the popularity spectrum.}

\begin{figure}
    \begin{subfigure}[b]{\linewidth}
        \includegraphics[width=0.7\linewidth]{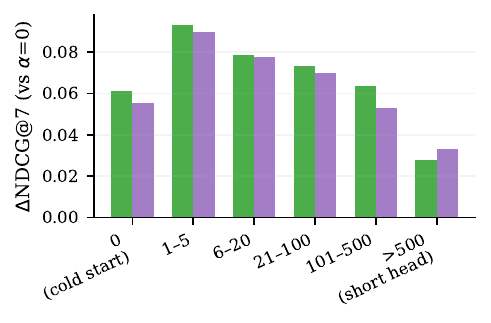}
        \caption{By query frequency}
        \label{fig:head-tail-query}
    \end{subfigure}\\
    \begin{subfigure}[b]{\linewidth}
        \includegraphics[width=0.7\linewidth]{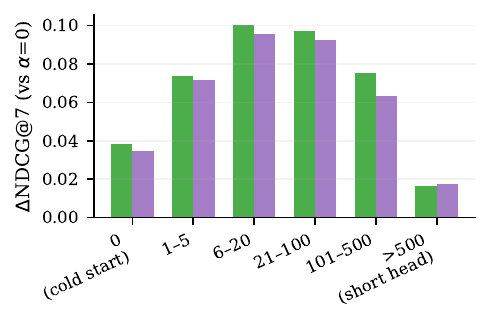}
        \caption{By item frequency}
        \label{fig:head-tail-item}
    \end{subfigure}
    \caption{NDCG@7 improvement over baseline ($\alpha{=}0$) by training frequency bin, sliced by (a) query frequency and (b) positive-item frequency. \protect\textcolor[HTML]{2ca02c}{\rule{6pt}{6pt}}~$\alpha{=}0.5$ (dual-sample) exceeds \protect\textcolor[HTML]{9467bd}{\rule{6pt}{6pt}}~$\alpha{=}1.0$ (features-only) on all but the short head queries ($>$500),
    where always seeing statistical features during training is beneficial.}
    \label{fig:head-tail-bars}
\end{figure}

To understand where dual-sample training provides the greatest benefit, we analyse NDCG@7 improvements over the $\alpha{=}0$ baseline across query- and item-frequency bins (Figure~\ref{fig:head-tail-bars}).

Both $\alpha{=}0.5$ and $\alpha{=}1.0$ improve over the baseline across all frequency ranges, but their relative behaviour differs systematically. On cold-start queries (frequency 0), $\alpha{=}0.5$ improves NDCG@7 by approximately $0.061$ compared with $0.055$ for $\alpha{=}1.0$ (Figure~\ref{fig:head-tail-query}), and this advantage persists throughout the long tail up to frequency 500. The trend reverses only on the short head ($>$ 500), where the features-only model performs slightly better.

Slicing by positive-item frequency (Figure~\ref{fig:head-tail-item}) reveals the same overall behaviour. Dual-sample training consistently outperforms features-only training across sparse and mid-frequency item ranges, with the largest gap appearing in the 101–500 frequency bin. Only on short-head items, where behavioural statistics are most reliable, does the features-only model perform comparably or slightly better.

\begin{figure}
    \begin{subfigure}[b]{\linewidth}
        \includegraphics[width=0.7\linewidth]{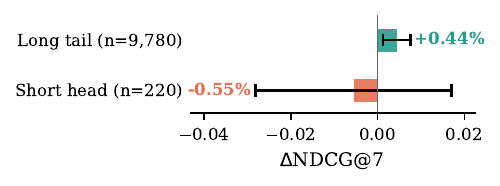}
        \caption{By query frequency}
        \label{fig:advantage-query}
    \end{subfigure}\\
    \begin{subfigure}[b]{\linewidth}
        \includegraphics[width=0.7\linewidth]{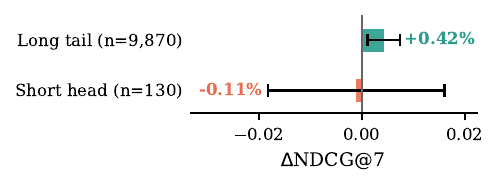}
        \caption{By item frequency}
        \label{fig:advantage-item}
    \end{subfigure}
    \caption{Advantage of dual-sample training ($\alpha{=}0.5$) over features-only training ($\alpha{=}1.0$), measured as the difference in NDCG@7 between the two models. Sessions are split into long tail ($\leq$500 training occurrences) and short head ($>$500). Error bars show 95\% CIs. On the long tail, $\alpha{=}0.5$ consistently outperforms $\alpha{=}1.0$;
    on the short head, the difference is not statistically significant.}
    \label{fig:head-tail-advantage}
\end{figure}

Figure~\ref{fig:head-tail-advantage} aggregates these effects across head and tail regions. On long-tail queries, dual-sample training improves NDCG@7 by +0.44\% (95\% CI: [0.12\%, 0.75\%]) relative to features-only training, whereas on the short head the point estimate is $-$0.55\% but not statistically  significant. Slicing by item frequency yields the same pattern: dual-sample training improves NDCG@7 by +0.42\% (95\% CI: [0.11\%, 0.74\%]) on long-tail items, while the short-head difference of $-$0.11\% is again not significant.

Taken together, these results show that dual-sample training primarily improves ranking in regimes where behavioural statistics are sparse or unreliable. When strong historical evidence is available, behavioural-feature-heavy training performs competitively or slightly better. In sparse and cold-start settings, however, explicitly training on paired QSS-present and QSS-removed views better preserves semantic and personalised relevance modelling.

Having established offline that dual-sample training improves robustness in sparse regimes while retaining most QSS-derived ranking gains, we next examine whether the same pattern appears under live production traffic.
 
\begin{figure}
    \begin{subfigure}[b]{\linewidth}
        \includegraphics[width=0.7\linewidth]{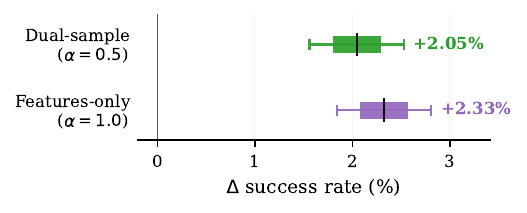}
        \caption{QSS-aware variants vs. $\alpha{=}0$ control}
        \label{fig:online-arm-control}
    \end{subfigure}\\
    \begin{subfigure}[b]{\linewidth}
        \includegraphics[width=0.7\linewidth]{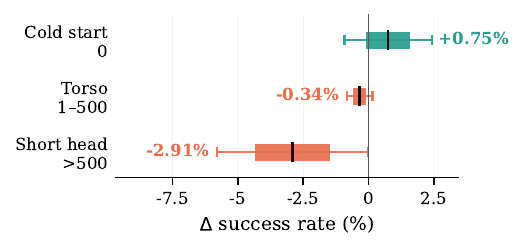}
        \caption{Dual-sample vs. features-only by item frequency}
        \label{fig:online-item-tiers}
    \end{subfigure}
    \caption{Online A/B/C test results on search success. Positive
    values indicate improvement over the $\alpha{=}0$ control cross-encoder in
    (a), and improvement of
    dual-sample training ($\alpha{=}0.5$) over features-only training
    ($\alpha{=}1.0$) in (b). Thick bars show $1\sigma$ intervals and thin lines
    show $2\sigma$ intervals, computed from per-user search success. In (b),
    item-frequency slices are cold start (0 training occurrences), torso
    (1--500), and short head ($>500$).}
    \label{fig:online-ab}
\end{figure}

\section{Online Evaluation}
\label{sec:online-test}

To evaluate whether the offline robustness patterns translate to production traffic, we conduct an online A/B/C test comparing three cross-encoder variants: a control model trained without QSS features
($\alpha{=}0$), dual-sample feature-dropout training ($\alpha{=}0.5$), and features-only
training ($\alpha{=}1.0$). Since the $\alpha{=}0$ model never saw QSS during training, QSS was also excluded from its prompt at serving time to avoid introducing noise (see Table~\ref{tab:offline-results}). The frontline metric is \emph{search success}: whether the
user has a successful interaction with a search result, such as streaming a
track, following an artist, or adding a podcast to their library. All analyses below are computed at the user level.

Figure~\ref{fig:online-ab}a shows that both QSS-aware variants significantly
improve search success relative to the $\alpha{=}0$ control cross-encoder. 
Dual-sample training improves search success by 2.05\%, while features-only
training improves it by 2.33\%. The difference between the two QSS-aware variants is not statistically significant: even the
$1\sigma$ confidence intervals overlap. Thus, exposing behavioural statistics through the prompt improves online ranking performance, but aggregate metrics alone do not distinguish which training objective better supports semantic generalisation.

To test whether the sparse-regime behaviour observed offline also appears online, we further slice traffic by item frequency 
(Figure~\ref{fig:online-ab}b). Each search is assigned to an item-frequency slice based on the maximum training-set frequency among the displayed items: cold start (0 occurrences), torso (1--500), and short head ($>500$). Search success is then computed independently within each slice.

This sliced analysis should be interpreted directionally rather than conclusively, since the online experiment was powered for aggregate search-success effects rather than fine-grained frequency slices. Dual-sample training shows a +0.75\% gain over features-only training on the cold-start slice, where displayed items are new to the training data and item-level behavioural statistics are unavailable. Although zero lies near the boundary of the $1\sigma$ interval, the pattern is consistent with the offline results and suggests that dual-sample training better preserves semantic and personalised relevance modelling when behavioural evidence is sparse. 
In contrast, the point estimates on torso and short-head items are
-0.34\% and -2.91\%, respectively, where behavioural statistics are most informative. This is consistent with the features-only model $\alpha{=}1.0$ relying more heavily on behavioural signals in high-frequency regimes.

Overall, the online results support the same qualitative pattern observed offline. Behavioural statistics substantially improve ranking performance, while dual-sample feature-dropout training appears to better preserve robustness on sparse and cold-start traffic. These findings suggest that explicitly controlling reliance on behavioural statistics may improve discoverability for new or infrequent content, while larger-scale online slice analyses remain an important direction for future work. We close by summarising what these results imply for exposing statistical features to production cross-encoders.

\subsection{Lessons and Limitations}
\label{sec:lessons-limitations}

\textbf{Signal-fusion lesson.} A strong behavioural statistic is not simply another prompt token: when it is highly predictive, the reranker can learn a brittle dependency that masks whether semantic and user-context evidence remain useful. Our results suggest a practical evaluation pattern for LLM rerankers: measure the same model with the feature present and under controlled removal, then inspect head, tail, and cold-start slices. In our setting, this exposes a failure hidden by aggregate full-prompt metrics and shows that paired feature removal retains most of the QSS benefit while improving robustness in sparse regimes.

\textbf{Operational lesson.} Paired training moves the robustness cost to training rather than serving. Each example yields matched QSS-present and QSS-removed views, increasing training computation, but deployment still uses one model with the same architecture and inference path. In our experiments, this trade-off preserves most of the QSS gain on feature-rich traffic while reducing degradation under QSS removal. For production rerankers, the practical advantage is direct: robustness to an unreliable feature does not require a separate fallback ranker or an architectural change.

\textbf{Evidence and scope.} Focusing on QSS in one personalised-search reranker gives a controlled view of the shortcut mechanism. Complete removal is intentionally a stress test rather than a reproduction of every partially missing, noisy, or stale feature scenario. The experiments treat user context as part of the overall prompt rather than isolating its independent contribution, and do not compare with a learned personalisation-aware fusion baseline. These choices sharpen the analysis of QSS reliance while leaving complementary questions about alternative fusion architectures open.

\textbf{Future directions.} The most immediate extension is to test graded perturbations that reflect realistic variation in feature availability, freshness, and noise, and to train with multiple correlated behavioural signals rather than removing one signal family completely. Other useful directions include feature-specific or adaptive mixing weights, learned gating or uncertainty-aware fusion, comparison with personalisation-aware rankers, and online experiments powered for cold-start and tail slices. Applying the same diagnostics across search and recommendation surfaces would help identify which robustness lessons transfer across domains and which depend on the feature or traffic regime.

\section{Conclusion}
\label{sec:conclusion}

This paper examined robust behavioural-signal fusion in an LLM cross-encoder for personalised search. The offline results establish the central failure mode: injecting QSS improves full-prompt NDCG@7 by 13.3\%, but QSS-always training degrades under the diagnostic QSS-removed condition. Deterministic paired feature removal preserves most of the QSS-derived gain while improving QSS-removed and long-tail performance, with its clearest advantages in sparse regimes.

The online A/B/C test complements these findings: both QSS-aware variants improve aggregate search success by approximately 2\%, confirming the production value of incorporating behavioural evidence into the reranker. In the cold-start slice, dual-sample training shows a directional gain consistent with the offline robustness pattern.

The practical conclusion is that strong behavioural prompt features should be evaluated both when available and under controlled removal, with explicit attention to head, tail, and cold-start traffic. Within the studied system, the results show that shortcut reliance can be diagnosed and reduced without changing the serving architecture. Extending this evaluation to richer feature families and other ranking surfaces is a natural next step.
 
\section*{Acknowledgments}

  \paragraph{Use of generative AI}
  Generative AI tools were used during the preparation of this work as coding assistants to help write analysis code and figure-generation scripts, and to polish portions of the manuscript. All outputs were reviewed, tested, and edited by the
  authors, who take full responsibility for the content of this paper.
\bibliographystyle{ACM-Reference-Format}
\bibliography{refs}

\end{document}